\documentclass[a4paper,11pt]{article}
\usepackage{jcappub} % for details on the use of the package, please
\usepackage{comment} % to be removed when we submit manuscript 

\usepackage[T1]{fontenc} % if needed

\usepackage{graphicx}
\usepackage{amsmath}	% required for `\cases' (yatex added)
\usepackage{siunitx}

\newcommand{\X}[2]{{#1}_\mathrm{#2}}
\newcommand{\T}[1]{\X{T}{#1}}
\newcommand{\kB}{\X{k}{B}}
\newcommand{\vc}{\X{v}{c}}
\newcommand{\vE}{\X{v}{E}}
\newcommand{\vEvec}{\X{\mathbf{v}}{E}}
\newcommand{\Abb}{A_\textrm{BB}}
\newcommand{\Tbb}{T_\textrm{BB}}
\newcommand{\Tbg}{T_\textrm{atm}}

\newcommand{\Trec}{T_\textrm{rec}}
\newcommand{\Tin}{T_\textrm{in}}
\newcommand{\Aeff}{A_\textrm{eff}}
\newcommand{\dnu}{\Delta \nu}
\newcommand{\Pdet}{P_{\gamma'}}

\title{Search for hidden-photon cold dark matter using a K-band cryogenic receiver}

\author[a]{N. Tomita}
\author[b, 1]{S. Oguri,\note{Corresponding author.}}
\author[c]{Y. Inoue,}
\author[a, d]{M. Minowa,}
\author[e]{T. Nagasaki,}
\author[f]{J. Suzuki,}
\author[f]{O. Tajima,}

\affiliation[a]{Department of Physics, School of Science, The University of Tokyo, 7-3-1 Hongo, Bunkyo-ku, Tokyo 113-0033, Japan}
\affiliation[b]{Institute of Space and Astronautical Science (ISAS), Japan Aerospace Exploration Agency (JAXA), 3-1-1 Yoshinodai, Chuo-ku, Sagamihara, Kanagawa 252-5210, Japan}
\affiliation[c]{International Center for Elementary Particle Physics, The University of Tokyo, 7-3-1 Hongo, Bunkyo-ku, Tokyo 113-0033, Japan}
\affiliation[d]{Research Center for the Early Universe (RESCEU), School of Science, The University of Tokyo, 7-3-1 Hongo, Bunkyo-ku, Tokyo 113-0033, Japan}
\affiliation[e]{Institute of Particle and Nuclear Studies, High Energy Accelerator Research Organization (KEK), 1-1 Oho, Tsukuba, Ibaraki 305-0801, Japan}
\affiliation[f]{Department of Physics, Faculty of Science, Kyoto University, Kitashirakawa Oiwake-cho, Sakyo-ku, Kyoto, 606-8502, Japan}
\emailAdd{tomita@icepp.s.u-tokyo.ac.jp}
\emailAdd{oguri.shugo@jaxa.jp}
\emailAdd{berota@icepp.s.u-tokyo.ac.jp}
\emailAdd{minowa.phys@gmail.com}
\emailAdd{tnaga@post.kek.jp}
\emailAdd{suzuki.junya.4r@kyoto-u.ac.jp}
\emailAdd{tajima.osamu.8a@kyoto-u.ac.jp}

\abstract{
We search for hidden-photon cold dark matter (HP-CDM) using a spectroscopic system in a K-band frequency range.
Our system comprises a planar metal plate and cryogenic receiver.
This is the first time a cryogenic receiver has been used in the search for HP-CDM.
Such use helps reduce thermal noise.
We recorded data for 9.3 hours using an effective aperture area of 14.8\,cm$^2$.
No signal was found in the data.
We set upper limits for the parameter of mixing between the photon and HP-CDM in the mass range from 115.79 to 115.85 $\mu$eV, $\chi < 1.8$--$\num{4.3e-10}$, at a 95\% confidence level.
This is the most stringent upper limit obtained to date in the considered mass range.

\vspace{5mm}
}

\begin{document}
\maketitle
\flushbottom

\section{Introduction}
\label{sec:introduction}

Various astronomical observations support the existence of cold dark matter (CDM).
In particular, they show that CDM localizes in galaxy halos as non-relativistic matter.
We understand that the gravity of galaxies is dominated by CDM.
However, we do not know whether CDM interacts with standard-model particles except through gravity.
A weakly interacting massive particle is a popular candidate of CDM,
and it is expected to have heavy mass (more than a GeV in typical models).
Many experiments have made efforts towards its detection.
However, there has not yet been any conclusive result \cite{DarkSide2018, DEAP2018, SuperCDMS2017, LUX2017, XENON1T2016, PandaXII2017}.
In recent years, another theoretical candidate, the weakly interacting slim particle, has been suggested.
The weakly interacting slim particle is expected to have a very small mass \cite{Arias2012, Jaeckel:2013ija}.
The hidden photon (HP) is a candidate weakly interacting slim particle and is
a $\rm{U}(1)$ gauge boson that kinetically mixes with ordinary photons \cite{JAECKEL2008509}.
The effective Lagrangian in this model is
\begin{equation}
\mathcal{L} =
 - \frac{1}{4} F_{\mu\nu} F^{\mu\nu}
 - \frac{1}{4} \tilde{X}_{\mu\nu} \tilde{X}^{\mu\nu}
 - \frac{1}{2} \chi F_{\mu\nu} \tilde{X}^{\mu\nu}
 + \frac{1}{2} m_{\gamma'}^{2} \tilde{X}_{\mu} \tilde{X}^{\mu},
 \label{eq:lagrangian}
\end{equation}
where $F_{\mu\nu}$ is the field strength of the electromagnetic field,
$\tilde{X}_{\mu\nu}$ is the field strength of the HP field ($\tilde{X}^\mu$),
$m_{\gamma '}$ is the mass of the HP,
and $\chi$ is the coupling constant of kinetic mixing.
The HP as CDM (hereafter HP-CDM) 
has been searched for via the kinetic mixing term (i.e., the third term in the above equation).
Thus far, experiments conducted using cavities have given strong upper limits at a mass below 20~$\mu$eV \cite{PhysRevLett.105.171801}.

A methodology of searching for HP-CDM in the higher mass region has been suggested~\cite{Horns:2012jf}.
Kinetic mixing converts HP-CDM into the electromagnetic waves (i.e., photons) at the surface of a metal plate.
The frequency of the conversion photon ($\nu_0$) represents the mass of the HP ($m_{\gamma '}$),
\begin{equation}
 h \nu_0 = m_{\gamma '} c^2, 
\end{equation}
where $h$ and $c$ are respectively the Planck constant and speed of light.
The original proposal suggested the use of a spherical metal plate for concentrating conversion photons whereas a recently published paper suggested combining a planar plate and parabolic mirror~\cite{Suzuki:2015vka}.

In experiments using a plate, the mixing angle $\chi$ is determined 
as described in ref.~\cite{Horns:2012jf},
\begin{equation}
\chi = 4.5 \times 10^{-14} 
	\left( \frac{\Pdet}{10^{-23}\,{\rm W}} \right)^{1/2} 
	\left( \frac{\rm 1\,m^{2}}{A_{\rm eff}} \right)^{1/2} 	
	\left( \frac{\rm 0.3\,GeV/cm^{3}}{\rho_{\rm CDM, halo}} \right)^{1/2} 	
	\left( \frac{\sqrt{2/3}}{\alpha} \right), \label{eq:mixing_angle}
\end{equation}
where $\Pdet$ is the measured power of the conversion photon, 
$A_{\rm eff}$ is the effective area of the antenna,
$\rho_{\rm CDM, halo}$ is the CDM density around the Solar System
($\SI{0.39 +- 0.03}{GeV/cm^3}$~\cite{1475-7516-2010-08-004}),
and $\alpha$ is a factor determined by the angular distribution of the HP-CDM field relative to the sensitive polarization axis.
For experiments using a single-polarization detector, we chose $\alpha = \sqrt{1/3}$ in the case of a random distribution~\cite{Arias2012}.
The sensitivity of the HP-CDM search depends on $A_{\rm eff}$ and a detection limit of power, $\Pdet = NEP/\sqrt{t}$, 
where $NEP$ is the noise equivalent power of the system while $t$ is the data integration time.

In this paper, we search for the HP-CDM using a planar plate and a receiver.
The detection principle is the same as that in ref.~\cite{Suzuki:2015vka}.
We focus on improving the $NEP$ using a cryogenic receiver.
We describe the experimental setup in section~\ref{sec:experimental-setup}, respectively describe
the data set and calibrations in sections~\ref{sec:dataset} and \ref{sec:calib}, present
the analysis method in section~\ref{sec:analysis}, give 
search results in section~\ref{sec:results}, and present
conclusions in section~\ref{sec:conclusion}.

\begin{figure}[tb]
 \centering
 \includegraphics[width=0.8\linewidth]{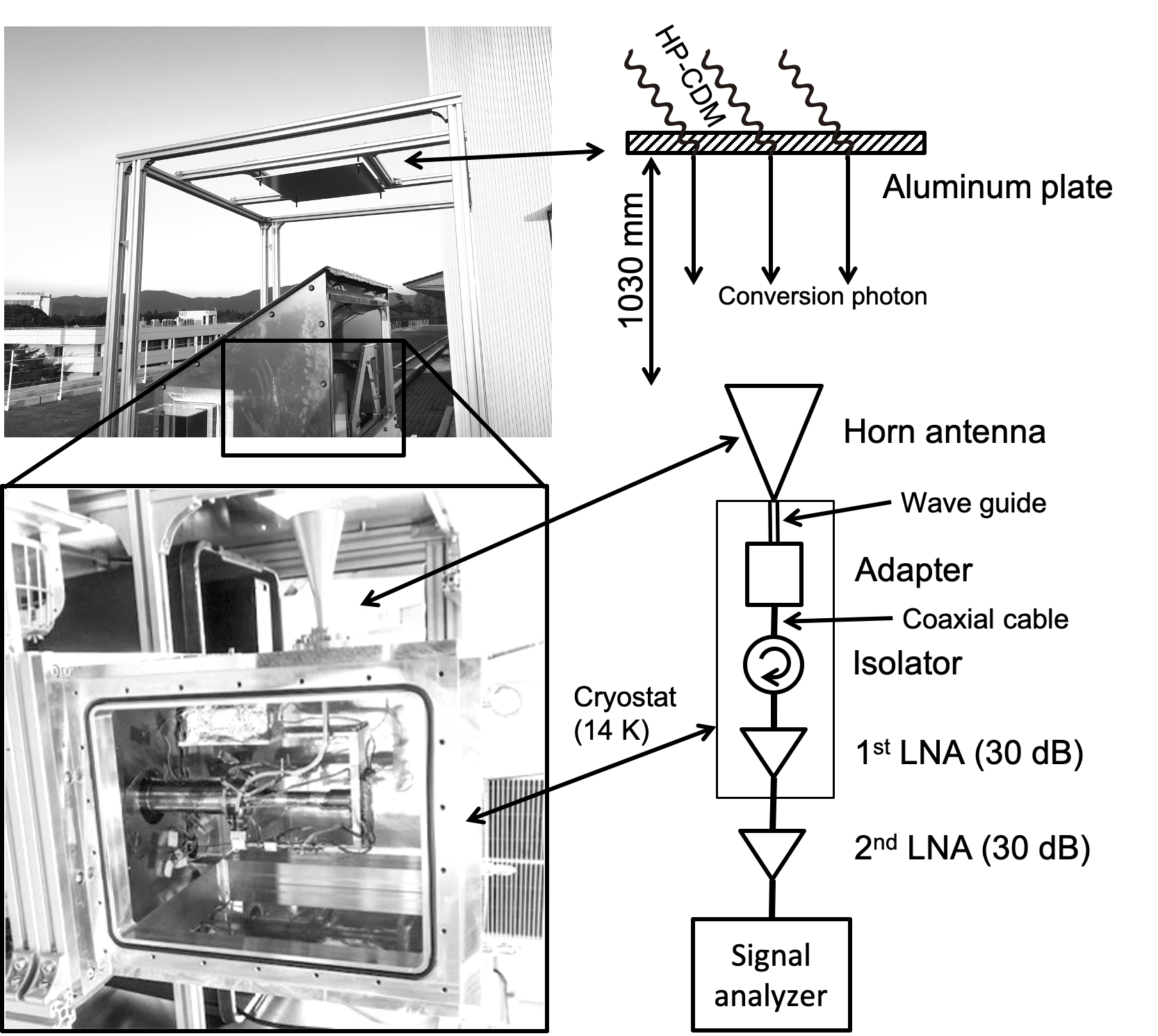}
 \caption{Overview of the experimental setup.}
 \label{fig:setup}
\end{figure}

\section{Experimental setup}
\label{sec:experimental-setup}

\begin{table}[tb]
 \centering
 \caption{Information on parts shown in Figure~\ref{fig:setup}.}
 \label{tab:config}
 \begin{tabular}{ll}
  \hline
  Aluminum plate & 500 mm $\times$ 500 mm $\times$ 6 mm, A5052-H112 \\
  Horn antenna   & Millitech, SGH-42-SC000, circular aperture ($\phi$\,58\,mm) \\
  Adapter  & SAGE Millimeter, SWC-422F-R1\\
  Isolator       & DiTom Microwave, D3I2030 \\
  $1^\textrm{st}$ LNA    & Low Noise Factory, LNF-LNC15\_29A \\
  $2^\textrm{nd}$ LNA    & Aldetec, ALM-1826S210 \\
  Signal analyzer & Agilent Technologies, N9010A (option 532) \\
  \hline
 \end{tabular}
\end{table}

\begin{table}
 \caption{Configuration parameters of the signal analyzer}
 \label{tab:speana}
 \centering
 \begin{tabular}{ll}
  \hline
  Sweep mode            & Fast Fourier Transform \\
  Resolution band width & 1 kHz \\
  Window function       & Flat-top \\
  Number of points      & 10001 \\
  Scan range
  & 10\,MHz from 27.998\,GHz, 28.000\,GHz, 28.002\,GHz \\
  & (Each of them is divided into two regions \\
  & \hspace*{5mm} because the band width for FFT is 5\,MHz.) \\
  \hline
 \end{tabular}
\end{table}

Our system comprises an aluminum plate and a cryogenic receiver operating in the K-band as illustrated in Figure~\ref{fig:setup}.
HP-CDM converts the photon at the surface of the aluminum plate.
The direction of travel of the conversion photon is perpendicular to the plate within $\SI{0.06}{\degree}$~\cite{Suzuki:2015vka}.
The cryogenic receiver measures the intensity of an incoming signal as a function of the frequency~\cite{KUMODeS2016PIERS}.
We search for the conversion signal as a narrow peak in the measured spectrum.

The receiver comprises a horn antenna, a waveguide-to-coax adapter, an isolator, two low-noise amplifiers, and a signal analyzer.
These components are commercial parts as listed in Table~\ref{tab:config}.
The diameter of the aperture of the conical horn is $58$ mm.
The antenna has an effective area of $\Aeff = \SI{14.8}{cm^2}$ and angular resolution of $\theta_{1/2} = \SI{12.5}{\degree}$ (full width at half maximum).
The calibration is detailed in section~\ref{sec:calib}.
We use a cryocooler to maintain the first low-noise amplifier (LNA) at 14\,K.
The low-temperature condition improves the intrinsic noise of the receiver (i.e., receiver temperature, $\Trec$).
$\Trec$ is approximately 46~K, which is one order of magnitude better than that under the ambient condition ($\Trec \sim \SI{400}{\kelvin}$).
Configuration parameters of the signal analyzer are summarized in Table~\ref{tab:speana}.

\begin{figure}
 \centering
 \includegraphics[width=0.6\linewidth]{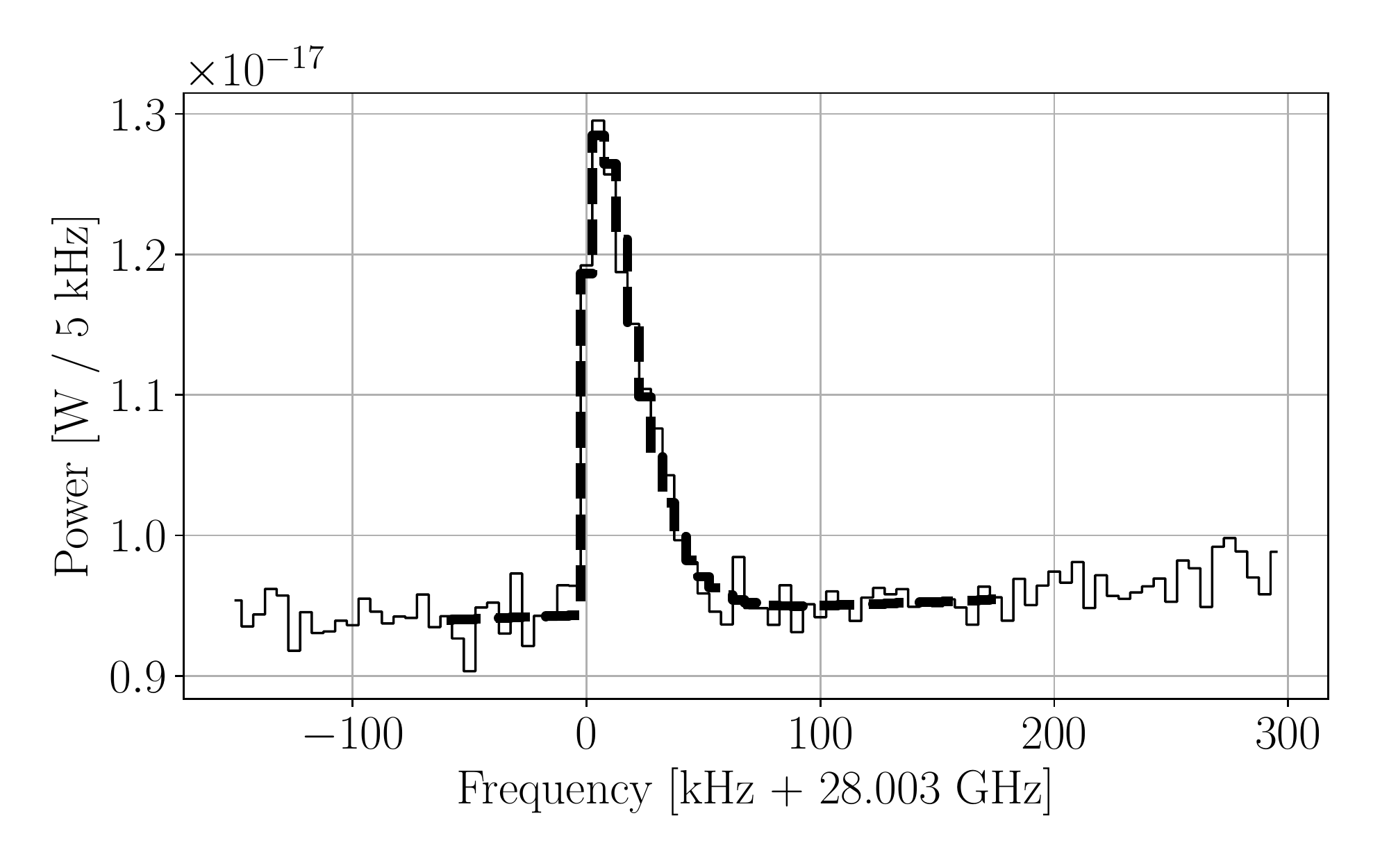}
 \caption{Spectrum obtained in a Monte Carlo simulation assuming $\chi = 2 \times 10^{-9}$ and $\nu_0 = \SI{28.003}{GHz}$ (thin solid line).
 We measure the signal power by fitting as described in the text of section~\ref{sec:analysis}.
 The fitting result for this simulation is overlaid as a thick dashed line.}
 \label{fig:eachfit}
\end{figure}

We assume a Maxwell--Boltzmann distribution (i.e., the standard halo model) for the velocity of the HP-CDM \cite{OHare2017}.
The distribution function of HP-CDM \cite[eq.~(15)]{PhysRevD.33.3495} is
\begin{equation}
 f({\bf v}) = \frac{1}{(\sqrt{\pi} \vc)^{3}} \exp{\left( - \frac{|{\bf v} + \vEvec |^2}{\vc^2} \right)},
\end{equation}
% \begin{equation}
%  \vc = \frac{\sqrt{\pi}}{2} \left< v \right>,
% \end{equation}
where ${\bf v}$ is the velocity of the HP-CDM,
%$\left< v \right>$ is the mean speed of the HP-CDM,
$\vc$ is the circular rotation speed of our Galaxy,
and $\vEvec$ is the velocity of the Earth in the frame of our Galaxy.
We can take $\left| \vEvec \right| \sim \vc$,
and the value of $\vc \sim \SI{220}{km / s}$ is adopted by many dark matter search experiments
\cite{DarkSide2018, LUX2017, PandaXII2017, Ahmed:2010hw, Aprile:2018dbl, Akerib:2018lyp}.
Because the relativistic energy is conserved when HP-CDM is converted to the photon,
the velocity of HP-CDM ($v \equiv |{\bf v}|$) is transformed to the frequency of the photon ($\nu$) according to
\begin{equation}
 h \nu = \frac{h \nu_0}{\sqrt{1 - (v/c)^2}}
  \quad
  \left( h \nu_0 \equiv m_{\gamma'} c^2 \right)
  \quad \Leftrightarrow \quad
  v = c \sqrt{1 - \left(\frac{\nu_0}{\nu}\right)^2}.
\label{eq:v_conv}
\end{equation}
Figure~\ref{fig:eachfit} shows the simulated spectrum assuming $m_{\gamma '} = \SI{115.81}{\mu eV}$ ($\nu_0 = \SI{28.003}{GHz}$) with the background being a linearly shaped offset.
The width of the signal over the peak frequency ($\dnu / \nu_0$) is approximately $10^{-6}$.

\section{Data set}
\label{sec:dataset}

We recorded data on the rooftop of a building ($\ang{36;8;53}$~north, $\ang{140;4;26}$~east, altitude of 40\,m above sea level).
We set the line of sight of the antenna to the zenith and set the plate above the antenna.
An advantage of the system configuration is the reduction of the flux of background radiation striking the antenna.
This is because that extra thermal radiation from an outside the field of the plate are dominated by the atmospheric radiation.
The power of thermal radiation passing into the receiver ($\approx 130$\,K) is lower than that in the case of ambient radiation ($\approx 300$\,K).

We performed the HP-CDM search from 6~p.m.~on~the 27th to 6~a.m.~on~the 28th of October~2016.
The measured frequency range is from 27.998 to 28.012\,GHz, corresponding to a mass range of 115.79 to 115.85\,$\mu$eV.
There are three scan ranges:
27.998--28.008,
28.000--28.010,
and
28.002--28.012\,GHz.
Each scan range is divided into two regions because the range for the Fourier transformation is 5\,MHz.
We therefore have six data regions in total for the analysis.
Scanning the six regions takes 11\,seconds.
We repeated this series of scans for a few hours.
Each repeated data set is named ``run''.
There are seven runs in total.
We calibrated the receiver gain and noise temperature at the beginning of each run and at the end of the last run.

\section{Calibration}
\label{sec:calib}

Measured signals for each frequency bin are modeled as
\begin{equation}
 S(\nu; \Tin) = G(\nu) \kB \left[ \Tin(\nu) + \Trec(\nu) \right] \dnu,
  \label{eq:measured_power} 
\end{equation}
where $G(\nu)$ is the receiver gain at frequency $\nu$,
$\kB$ is the Boltzmann constant,
$\Tin(\nu)$ is radiation passing into the receiver with Rayleigh-Jeans unit of Kelvin (K),
and $\dnu$ is the frequency bin width.
The calibration uses a blackbody source (ECCOSORB CV-3 from E\&C Engineering), which fully covers the aperture of the horn antenna.
For each calibration, we recorded two series of data using the blackbody at an ambient temperature ($\T{ambient} \approx \SI{300}{K}$) and liquid-nitrogen temperature ($\T{LN_2} = \SI{77}{K}$).
Using these two data series, we obtain $G(\nu)$ and $\Trec(\nu)$ as
\begin{equation}
 G(\nu) = \frac{S(\nu; \T{ambient}) - S(\nu; \T{LN_2})}{\kB (\T{ambient} - \T{LN_2}) \dnu},
\end{equation}
\begin{equation}
 \Trec(\nu) = \frac{S(\nu; \T{LN_2})}{G(\nu) \kB \dnu} - \T{LN_2}.
\end{equation}
Table~\ref{tab:g_trec_cal} gives results for each calibration.
The average gain and noise temperature are approximately $\SI{66}{dB}$ and $\SI{46}{K}$, respectively.
Their time variations are less than the uncertainty in the CDM density (7.7\%).
Through the interpolation of calibration results before and after each run, 
we measure the power entering the receiver in each frequency bin: $S(\nu) / ( G(\nu) \dnu) - \kB \Trec$.

\begin{table}[b]
\centering
\caption{Receiver gains and receiver temperatures measured soon before each run and after the last run.
Here, $\left< G \right>$ and $\left< \Trec \right>$ are the average gain and receiver temperature in the six frequency regions.}
\label{tab:g_trec_cal}
\begin{tabular}{rccc}
 \hline \hline
 \multicolumn{1}{c}{Date time} &
 $\left< G \right>$ [dB] & $\left< \Trec \right>$ [K] & $\T{ambient}$ [K] \\
 \hline
October 27th 18:36 & 65.67 $\pm$ 0.02 & 48.2 $\pm$ 0.3 & 290.3 \\
             19:40 & 65.78 $\pm$ 0.04 & 46.4 $\pm$ 0.3 & 289.5 \\
             21:36 & 66.01 $\pm$ 0.03 & 45.9 $\pm$ 0.3 & 288.4 \\
             23:43 & 66.13 $\pm$ 0.02 & 45.7 $\pm$ 0.3 & 287.3 \\
October 28th 02:05 & 66.57 $\pm$ 0.02 & 45.4 $\pm$ 0.4 & 284.8 \\
             03:05 & 66.78 $\pm$ 0.02 & 45.2 $\pm$ 0.3 & 283.6 \\
             04:05 & 66.81 $\pm$ 0.02 & 45.9 $\pm$ 0.4 & 283.8 \\
             05:32 & 66.93 $\pm$ 0.03 & 45.2 $\pm$ 0.5 & 283.8 \\
 \hline
\end{tabular}
\end{table}

\begin{figure}[b]
 \centering
 \includegraphics[width=0.7\linewidth]{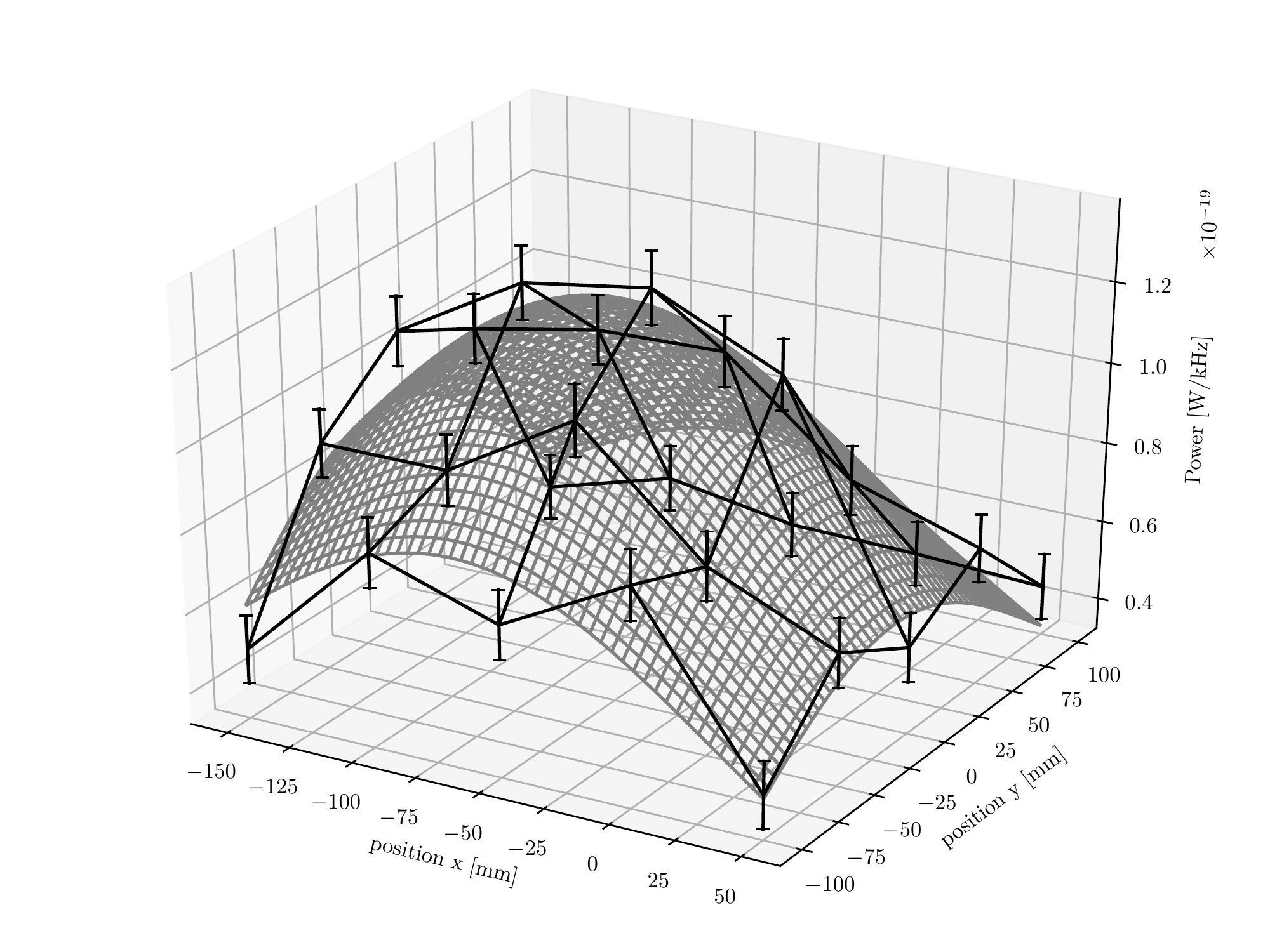}
 \caption{Measured powers with changing blackbody location.
 We obtain the beam shape and power at the beam center through fitting.
 We take these calibration data for each frequency region.}
 \label{fig:cal_beam_shape}
\end{figure}

The effective area ($\Aeff$) and beam width ($\theta_{1/2}$) of the horn antenna are calibrated using a round-shaped blackbody made by the ECCOSORB AN-72 (E\&C Engineering).
The blackbody has a diameter of 6\,cm; i.e., its area ($\Abb$) is $\SI{28.3}{cm^2}$.
%$\SI{28.27 \pm 0.94}{cm^2}$
We removed the aluminum plate during this calibration.
Changing the location of the blackbody in the ambient, we measured powers at each location.
In each step, we also measured the atmospheric radiation without the blackbody.
We recorded calibration data on October 24th and 26th.
The ambient temperatures were 299\,K on the 24th and 302\,K on the 26th.
We measured $G(\nu)$ and $\Trec(\nu)$ before and after calibration on each day.
The distribution of the measured power is modeled with a Gaussian beam shape \cite{Goldsmith1992},
\begin{equation*}
 \frac{\Aeff \cos(\X{\theta}{BB}) \cdot \Abb}{L^2} \frac{\nu^2}{c^2}
 \times
 \exp \left[ - 2 \frac{(x - x_0)^2 + (y - y_0)^2}{w^2} \right]
 \times
 \kB (\Tbb - \Tbg)
\end{equation*}
\begin{equation}
 \left(
  \tan(\X{\theta}{BB}) = \frac{\sqrt{(x - x_0)^2 + (y - y_0)^2}}{L}
  , \qquad
  w = \frac{2 L}{\sqrt{2\log 2}} \tan(\X{\theta}{1/2} / 2)
 \right),\label{eq:gaussian_beam}
\end{equation}
where $(x, y)$ is the location of the blackbody,
$(x_0, y_0)$ is the line of sight,
$\X{\theta}{BB}$ is the direction of the blackbody with respect to the line of sight,
$w$ is the beam size on the plane of the blackbody,
$\Tbb$ is the blackbody temperature in the ambient,
$\Tbg$ is the atmospheric radiation temperature,
and $L$ is the distance from the horn antenna to the calibration source (1030\,mm).

For each calibration, we obtain $\Aeff$, $\theta_{1/2}$, and $(x_0, y_0)$ by fitting with eq.~\eqref{eq:gaussian_beam} as shown in Figure~\ref{fig:cal_beam_shape}.
Fitting results for each calibration are summarized in Table~\ref{tab:beamfit}.
By taking a weighted average, we obtain $\Aeff = \SI{14.8 +- 0.9}{cm^2}$ and $\theta_{1/2} = \SI{12.5 +- 0.2}{\degree}$.
We here assign errors as the square root of the quadrature sum of the deviation among calibrations, uncertainties for the gain, $\Trec$, $\Tbb$, $\Abb$, and $\Tbg$.

\begin{table}
 \caption{$\Aeff$ and $\theta_{1/2}$ for each frequency region on each day.}
 \label{tab:beamfit}
 \centering
 \begin{tabular}{cccccr}
\hline \hline
Date & Frequency [GHz] &
 \multicolumn{1}{c}{$\Aeff \, [\mathrm{cm^2}]$} &
 \multicolumn{1}{c}{$\theta_{1/2} \, [\mathrm{^\circ}]$} &
 \multicolumn{1}{c}{$x_0 \, [\mathrm{mm}]$} &
 \multicolumn{1}{c}{$y_0 \, [\mathrm{mm}]$} \\
\hline
October 24th & 27.998--28.003 & 14.6 $\pm$ 0.5 & 13.3 $\pm$ 0.6 & $-67.9 \pm 3.8$ & $-10.9 \pm 3.7$ \\
             & 28.000--28.005 & 14.4 $\pm$ 0.5 & 12.5 $\pm$ 0.5 & $-65.7 \pm 3.6$ & $  2.6 \pm 3.5$ \\
             & 28.002--28.007 & 14.8 $\pm$ 0.5 & 12.9 $\pm$ 0.5 & $-60.9 \pm 3.6$ & $-12.9 \pm 3.6$ \\
             & 28.003--28.008 & 15.0 $\pm$ 0.5 & 12.8 $\pm$ 0.5 & $-55.9 \pm 3.5$ & $-10.2 \pm 3.5$ \\
             & 28.005--28.010 & 15.9 $\pm$ 0.5 & 12.0 $\pm$ 0.4 & $-69.8 \pm 3.3$ & $-11.1 \pm 3.2$ \\
             & 28.007--28.012 & 15.5 $\pm$ 0.5 & 12.5 $\pm$ 0.5 & $-55.1 \pm 3.4$ & $-12.4 \pm 3.4$ \\
\hline
October 26th & 27.998--28.003 & 13.6 $\pm$ 0.6 & 13.5 $\pm$ 0.7 & $-64.6 \pm 4.9$ & $  1.1 \pm 4.7$ \\
             & 28.000--28.005 & 14.1 $\pm$ 0.6 & 12.6 $\pm$ 0.6 & $-77.3 \pm 4.7$ & $ -7.1 \pm 4.4$ \\
             & 28.002--28.007 & 14.1 $\pm$ 0.6 & 13.6 $\pm$ 0.7 & $-80.7 \pm 5.2$ & $  0.6 \pm 4.7$ \\
             & 28.003--28.008 & 15.5 $\pm$ 0.6 & 12.5 $\pm$ 0.6 & $-72.3 \pm 4.2$ & $  1.3 \pm 3.9$ \\
             & 28.005--28.010 & 14.6 $\pm$ 0.6 & 13.8 $\pm$ 0.7 & $-77.1 \pm 5.1$ & $ -8.3 \pm 4.7$ \\
             & 28.007--28.012 & 16.8 $\pm$ 0.7 & 11.1 $\pm$ 0.4 & $-71.3 \pm 3.5$ & $ -3.5 \pm 3.4$ \\
\hline
 \end{tabular}
\end{table}

\section{Analysis}
\label{sec:analysis}

%\subsection{Data Selection}

\begin{figure}
 \centering
 \includegraphics[width=0.45\linewidth]{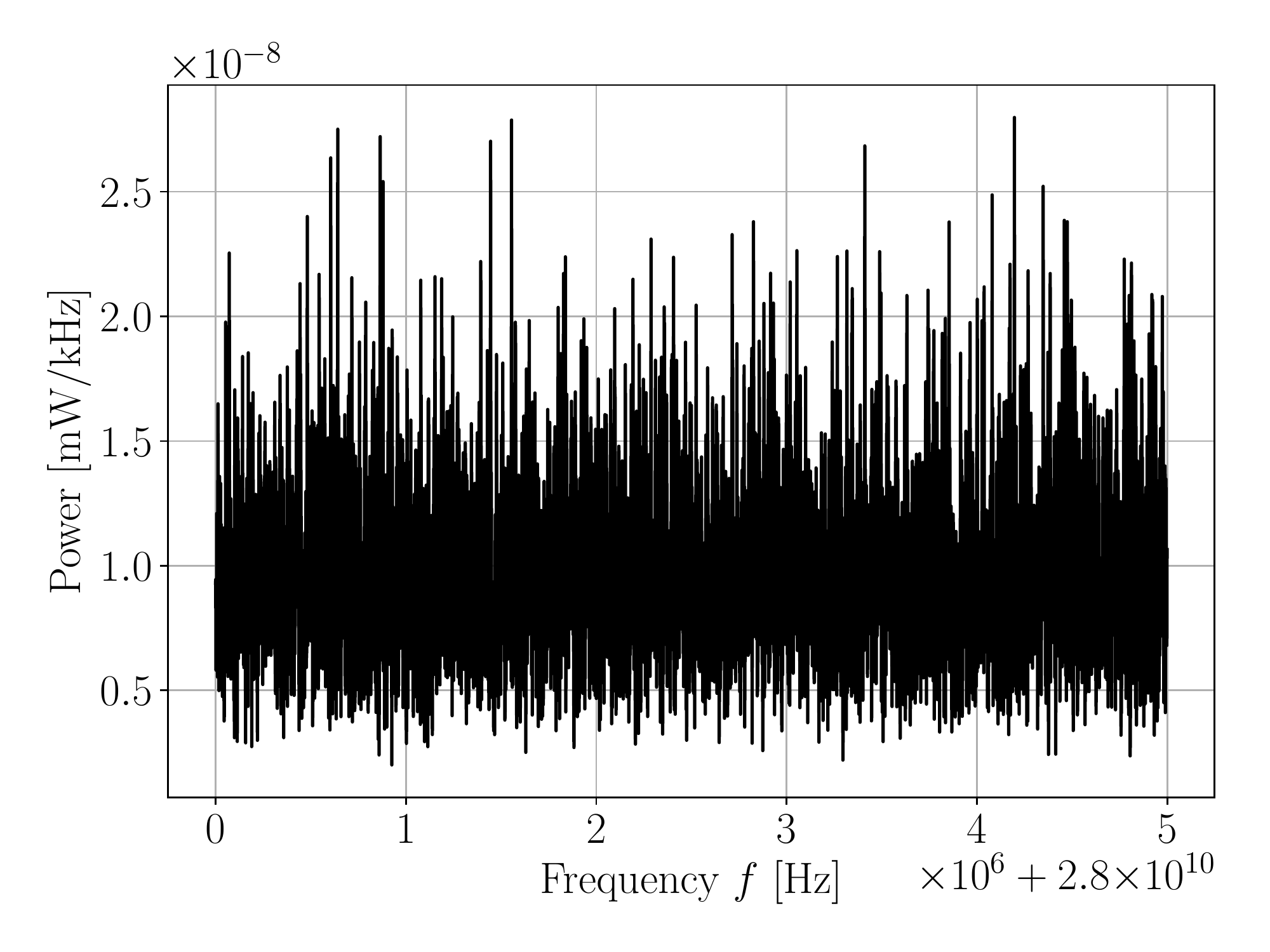}
 \includegraphics[width=0.45\linewidth]{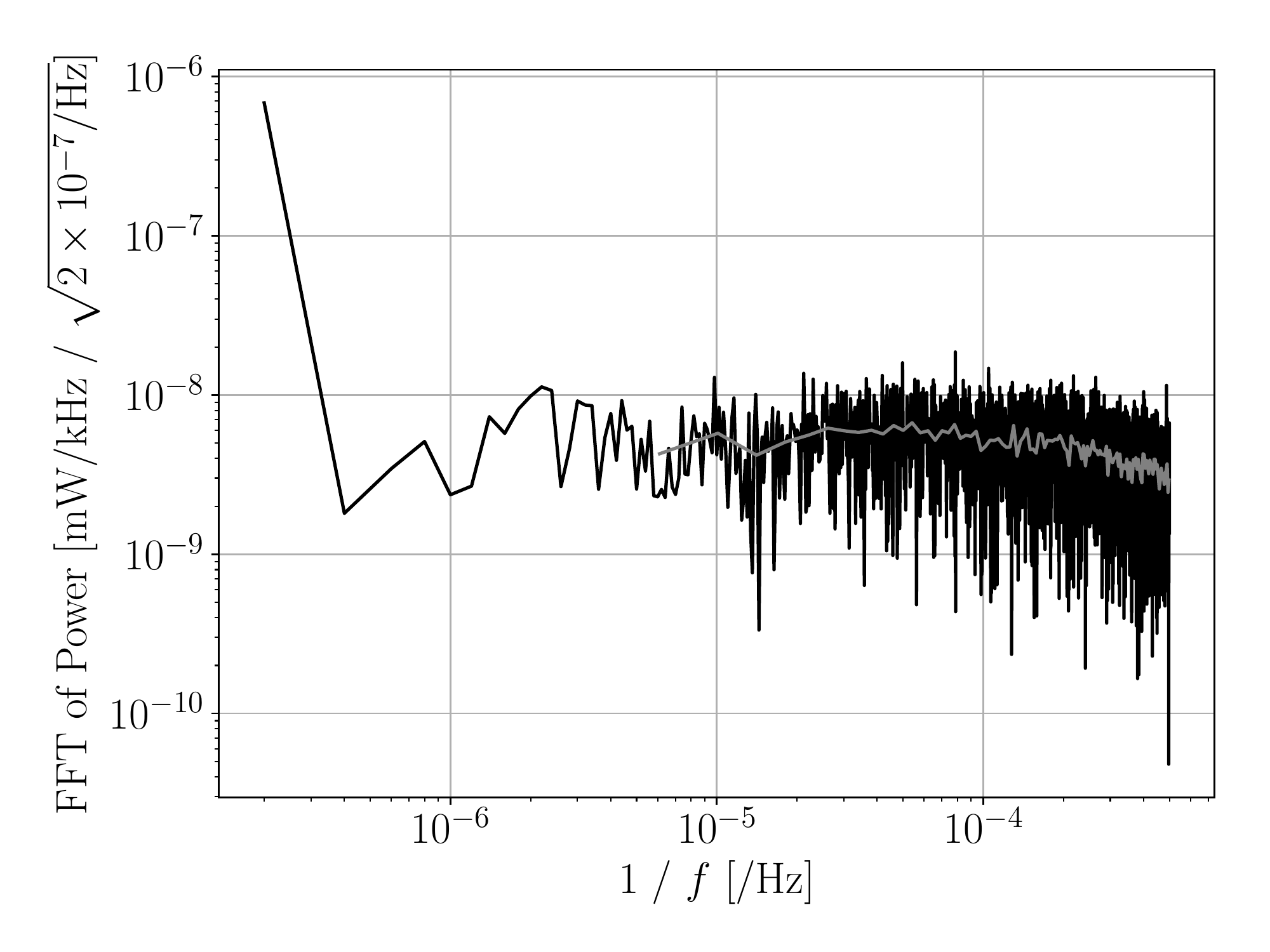}
 \includegraphics[width=0.9\linewidth]{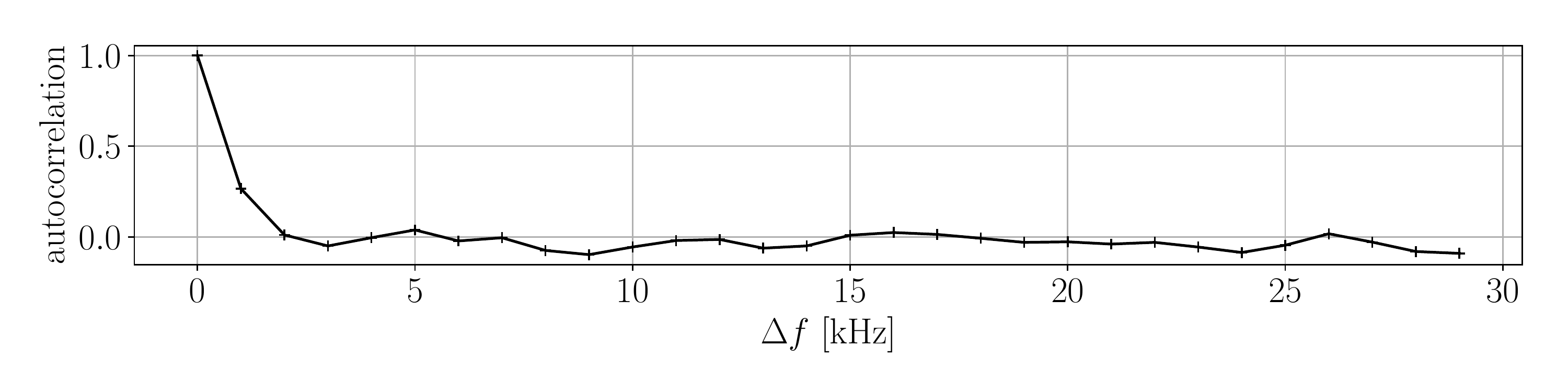}
 \caption{[top-left]
 Example of the power measured by the spectrometer.
 The bin width is 1\,kHz and there are 5000 samples in each frequency region.
 [top-right]
 The Fourier-transformed spectrum of measured powers.
 The gray line shows the average.
 White noise generates a flat structure in the region between $\mathcal{O}(\SI{1e-5}{})$ and $\mathcal{O}(\SI{1e-4}{})$\,/Hz.
 A cut-off due to correlation between adjacent bins is seen above $\mathcal{O}(\SI{1e-4}{})$\,/Hz.
 [bottom]
 Autocorrelation of the raw data shown at top-left.
 }
 \label{fig:rawdata}
\end{figure}

We search for the HP-CDM signal in frequency spectra for each run in each frequency region by fitting with a model function with changing $\nu_0$.
Before fitting, we re-bin the power spectra to eliminate systematic bias from correlation between adjacent frequency bins.
Figure~\ref{fig:rawdata} shows measured powers as a function of frequency, a Fourier-transformed spectrum, and the correlation among bins.
We find correlation among neighboring bins up to 3\,kHz.
We thus use a frequency bin width of 5\,kHz by summing for five samples.

We fit signal powers by changing $\nu_0$ in steps of 5~kHz.
A fitting function comprises the signal term and background terms.
We model the background shape with a first-order polynomial,
\begin{equation}
 f(\nu;P_{\gamma'},\nu_0,a,b) = P_{\gamma'} \, \left\{ F(\nu + \SI{2.5}{kHz};\nu_0) - F(\nu - \SI{2.5}{kHz};\nu_0) \right\} + a(\nu - \nu_0) + b,
\end{equation}
where $P_{\gamma'}$ is the power of the conversion photon, and cummulative distribution $F(\nu;\nu_0)$ is introduced to account for the effect of finite bin width.
Cumulative distribution of HP-CDM verocity $F_v(v)$ is
\begin{align}
 F_v(v)
 &= \int_{0}^{v} {\rm d}v^{\prime} \int^{4\pi} {\rm d}\Omega \, f({\bf v^{\prime}}) \, {v^{\prime}}^2 \nonumber \\
 \nonumber
 &= \frac{\vc}{2 \sqrt{\pi} \vE} \left\{ \exp \left[ - \left( \frac{v + \vE}{\vc} \right)^2 \right] - \exp \left[ - \left( \frac{v - \vE}{\vc} \right)^2 \right] \right\} \\
\label{eq:v_dist}
 & {} \qquad + \frac{1}{2} \left\{ {\rm erf} \left[ \frac{v - \vE}{\vc} \right] + {\rm erf} \left[ \frac{v + \vE}{\vc} \right] \right\}, \\
 {\rm erf}(x) &= \frac{2}{\sqrt{\pi}}  \int_{0}^{x} e^{-t^2} {\rm d}t.
 \label{eq:erf}
\end{align}
Plugging eq.~\eqref{eq:v_conv} into eq.~\eqref{eq:v_dist}, 
we obtain $F(\nu;\nu_0)$.
For each fit, we use a fitting range of 240~kHz (48~bins):
$(\nu0 - \SI{60}{kHz})$--$(\nu0 + \SI{180}{kHz})$.
We estimate the error in each bin as white noise
and assume it has the same size among bins.
We apply the Fourier transform to the frequency spectra as shown in Figure~\ref{fig:rawdata} [top-right].
We calculate the mean in the ``1 / frequency'' region between $\SI{6e-5}{/Hz}$ and $\SI{1e-4}{/Hz}$ and use it as the white-noise error.

%\subsection{Validations}

\begin{figure}
 \centering
 \includegraphics[width=0.6\linewidth]{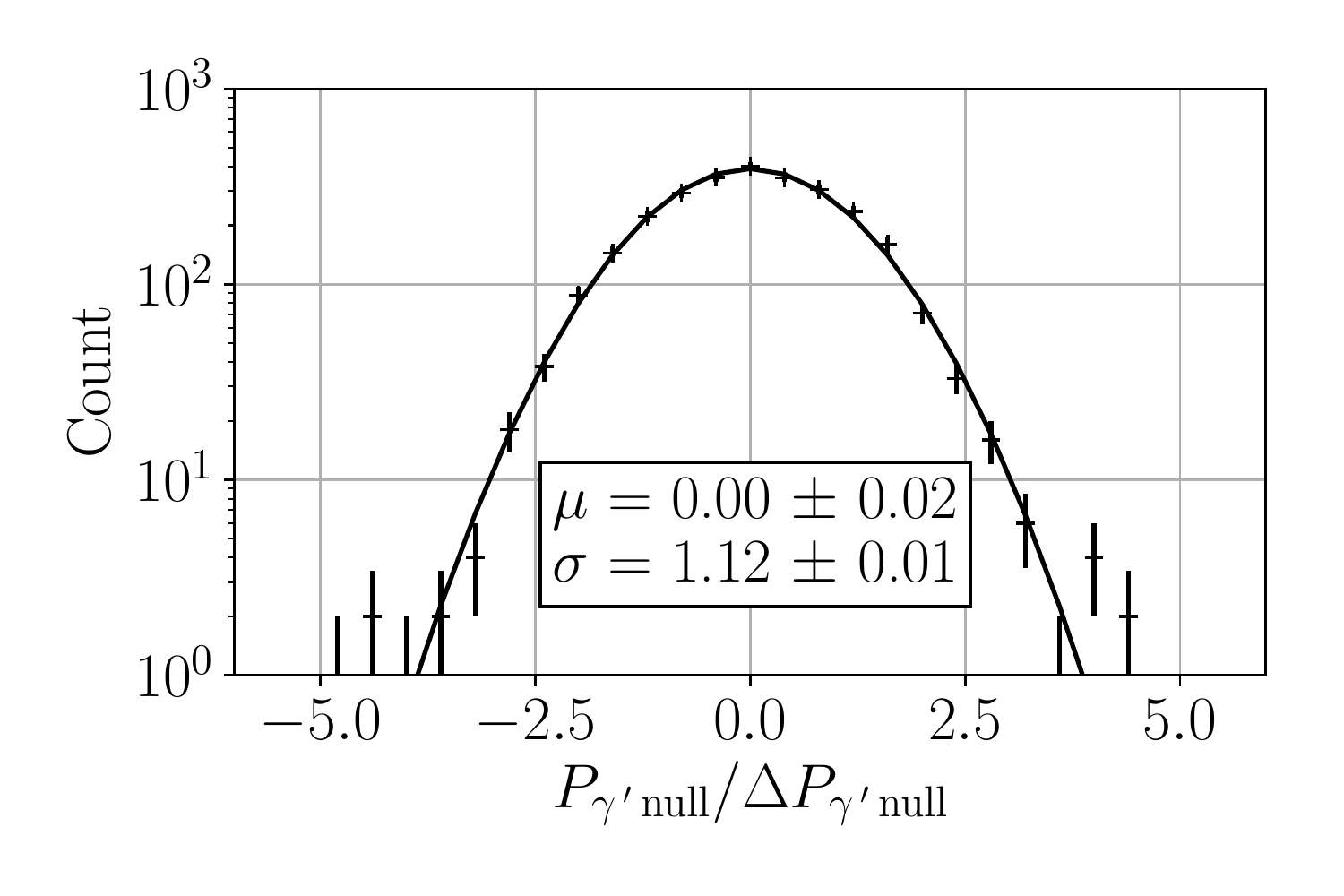}
 \caption{Distribution of $P_{\gamma'\,{\rm null}}/\Delta P_{\gamma'\,{\rm null}}$ for the null samples.
 Fitting results obtained with the Gaussian distribution are also shown.}
 \label{fig:distP_null}
\end{figure}

We validate the analysis procedure adopting a ``null sample'' method.
We randomly divide the calibrated data for each run into two data sets,
and we create subtracted data samples for each region (named null samples).
No signal is contained in a null sample because signals in each data set cancel out.
We check whether the mean value of the fitting results for null samples
($P_{\gamma'\,{\rm null}}$)
is consistent with a value of zero.
We also check for analysis bias between the fitting error
($\Delta P_{\gamma'\,{\rm null}}$) and a standard deviation of fitting results.
Figure~\ref{fig:distP_null} shows the distribution of $P_{\gamma'\,{\rm null}} / \Delta P_{\gamma'\,{\rm null}}$.
The distribution is fitted using a Gaussian curve, yielding a mean ($\mu = 0.00 \pm0.02$) and a standard deviation ($\sigma = 1.12 \pm 0.01$).
No bias is found for the mean value.
However, we understand that
the error used in the signal extraction should be corrected using a multiplication factor of 1.12.
The raw fitting error is deemed inaccurate because we only consider white noise in the fitting.
An imperfection of the background model may also have this effect.

%\subsection{Systematic uncertainties}

Systematic uncertainties are summarized in Table~\ref{tab:systematic_error}.
A small angle offset of the conversion photons ($< \SI{0.06}{\degree}$) slightly reduces the detection efficiency.
This efficiency loss is estimated from the beam width:
$1 - \exp\left[ -\SI{0.06}{\degree} / (\SI{12.25}{\degree} / 2) \cdot \log(2) \right]$ = 0.7\%.
Meanwhile, plate curvature, roughness, reflectivity, and the misalignment of the plate direction
result in a loss of signal.
In particular, plate curvature (i.e., deviation from flatness) results in non-negligible loss.
Plate curvature occurs because the aluminum plate is held at its edges and deforms under its own weight.
The deformation of the plate is measured using a ruler and
the curvature is estimated as $R = 62.5\,\mathrm{m}$ conservatively.
This gives uncertainty of $3.2\%$.
Systematic uncertainty in $\Aeff$ is determined by the calibration error described in section~\ref{sec:calib}.
We understand that the receiver gain is stable compared with the uncertainty in CDM density (7.7\%) as described in section~\ref{sec:calib}.
The frequency stability is examined using a vector network analyzer (VNA, Agilent, N5224A) as a microwave source.
The VNA generates a monochrome wave with a frequency of 28.003 GHz, and the signal analyzer measures the wave.
We confirm that the fluctuation of the frequency peak position is below 1~kHz
even if we change the temperature of the signal analyzer from 27.8 ${}^\circ$C to 40.3 ${}^\circ$C.
This effect corresponds to a tiny efficiency loss of 0.9\%.
A possible fitting bias due to the frequency bin is estimated using the Monte Carlo simulation: $4.8\%$.
The total systematic uncertainty is assigned to be 11.3\% for signal detection.

\begin{table}[tbp]
\centering
\caption{
Systematic uncertainties of the HP-CDM signal.}
\label{tab:systematic_error}
\begin{tabular}{@{\hspace{3mm}} l @{\hspace{6mm}} r @{\hspace{3mm}}}
\hline
\multicolumn{1}{c}{Source} & \multicolumn{1}{c}{\%} \\
\hline
Angular dispersion of HP-CDM & 0.7 \\ % $6.7 \times 10^{-1}$ \\
Alignment, flatness, roughness, and reflectivity of the aluminum plate & 3.2 \\ % sqrt( 0.3**2 + 3.2**2 + 9.6*10**-8 +0.079**2)
Effective area of antenna ($A_{\rm eff}$) & 5.8 \\
Responsivity of the receiver & 7.7 \\
Frequency response & 0.9 \\ % Possible instability of frequency response 0.90
Frequency bin & 4.8 \\ % possible fit bias including binning effect
%Statistical error estimation & 2.3 \\
\hline
Total & 11.3 \\
\hline
\end{tabular}
\end{table}

\section{Results}
\label{sec:results}

\begin{figure}[tb]
 \centering
 \includegraphics[width=0.8\linewidth]{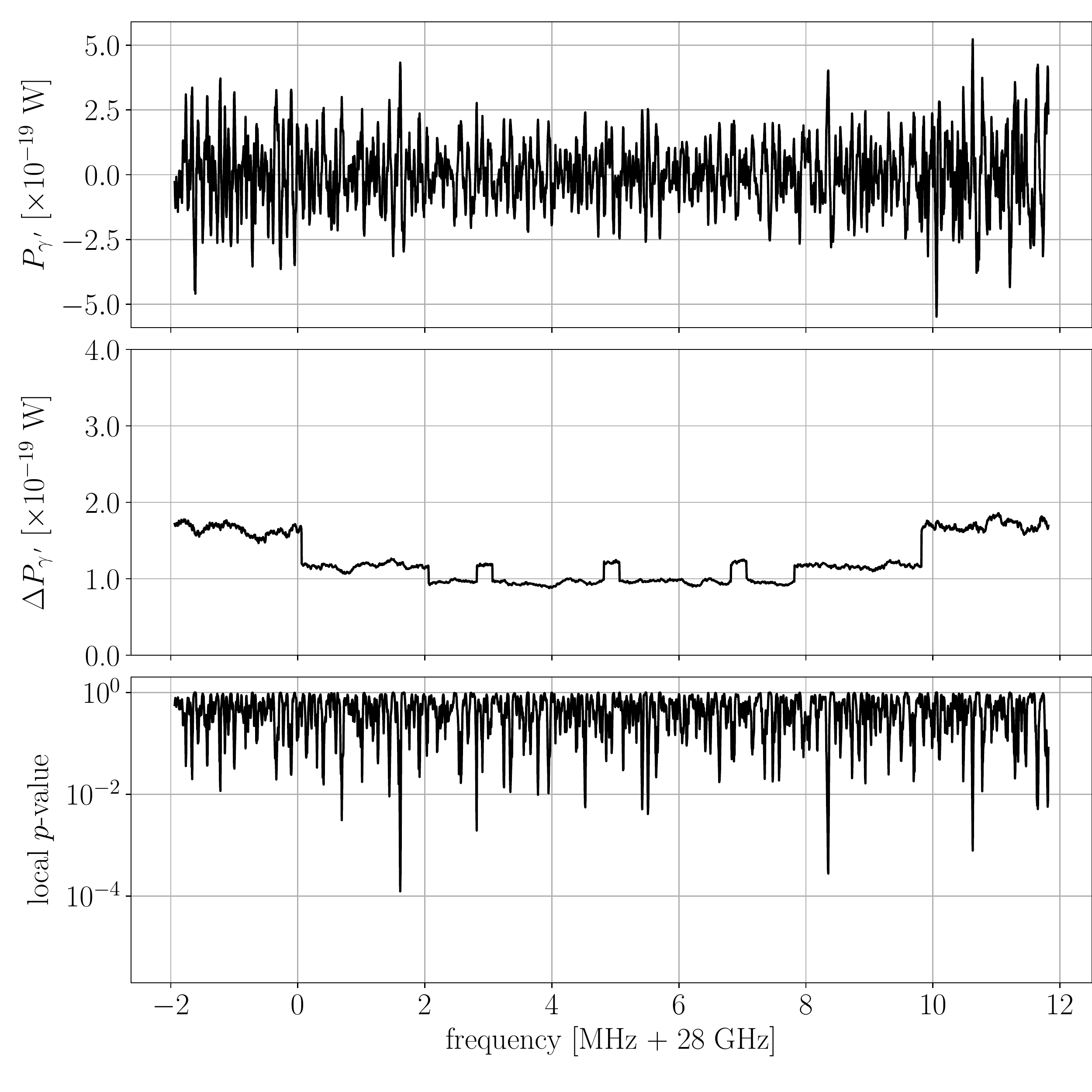}
 \caption{
 Extracted powers for the HP-CDM signal as a function of frequency [top],
 statistical errors [middle], and local p-values [bottom].}
 \label{fig:power_data}
\end{figure}

Figure~\ref{fig:power_data} shows the extracted power ($P_{\gamma'}$) and statistical error ($\Delta P_{\gamma'}$) as functions of frequency.
Local $p$-values, $(1 - {\rm erf}(P_{\gamma'}/\Delta P_{\gamma'} / \sqrt{2}))/2$, are also shown.
We apply the correction for statistical error (i.e., a multiplication of 1.12) as described in the previous section.
The minimum local $p$-value in 2752 frequency bins is $p_{\rm min} = \SI{1.24e-4}{}$ ($\SI{3.7}{\sigma}$).
Adopting the methodology described in ref.~\cite{PhysRevD.97.123006}, we account for the look elsewhere effect.
We determine the number of independent frequency windows ($\num{1.0e3}$) by conducting Monte Carlo simulations.
The probability of exceeding $\X{p}{min}$ in any frequency bin is estimated as
\begin{equation}
 1 - (1 - p_{\rm min})^{\num{1.0e3}} = 0.12 \quad (1.2 \sigma).
\end{equation}
We do not find any significant excess of the HP-CDM signal from zero.

\begin{figure}[tb]
 \centering
 \includegraphics[width=0.8\linewidth]{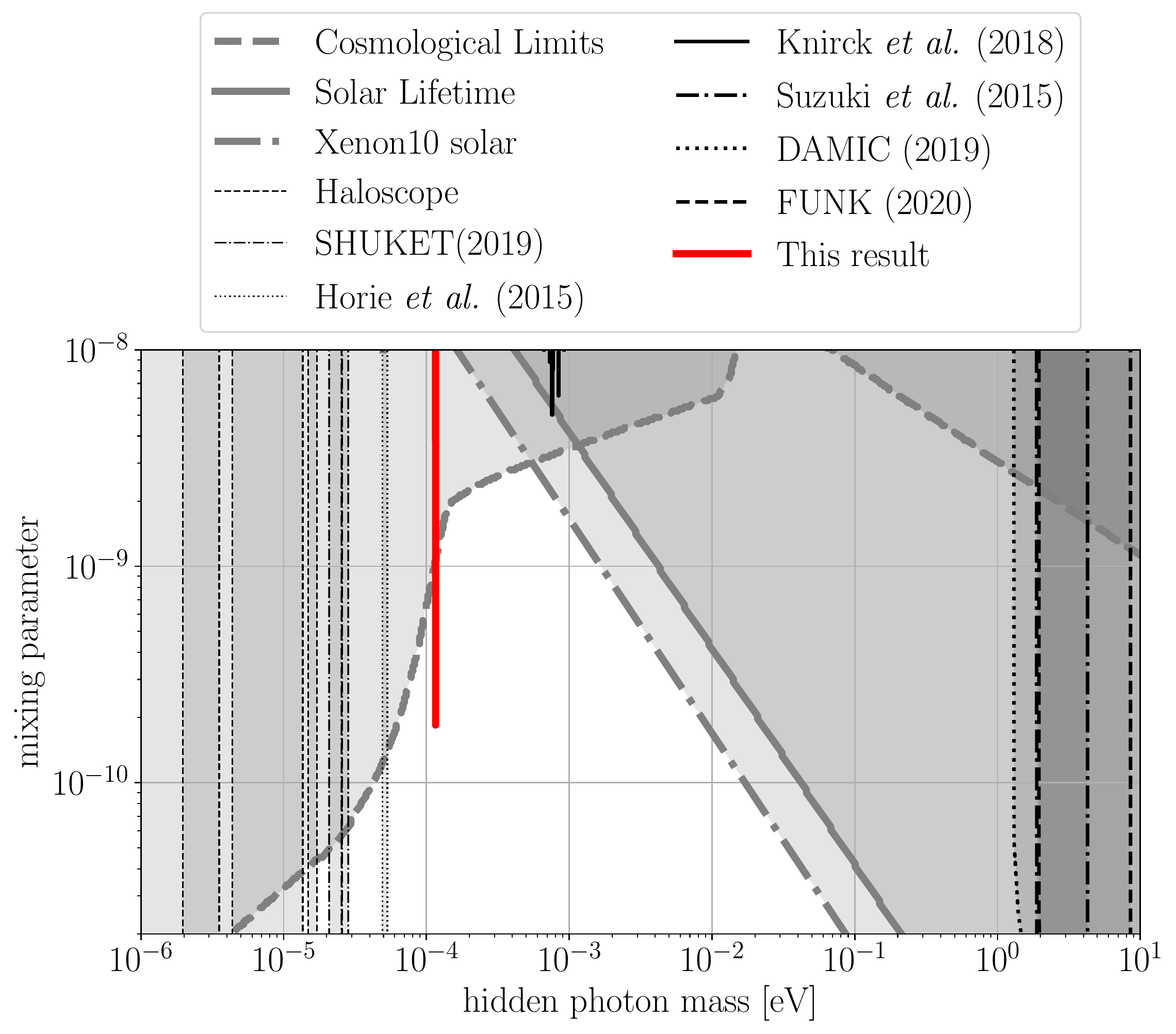}
 \includegraphics[width=0.8\linewidth]{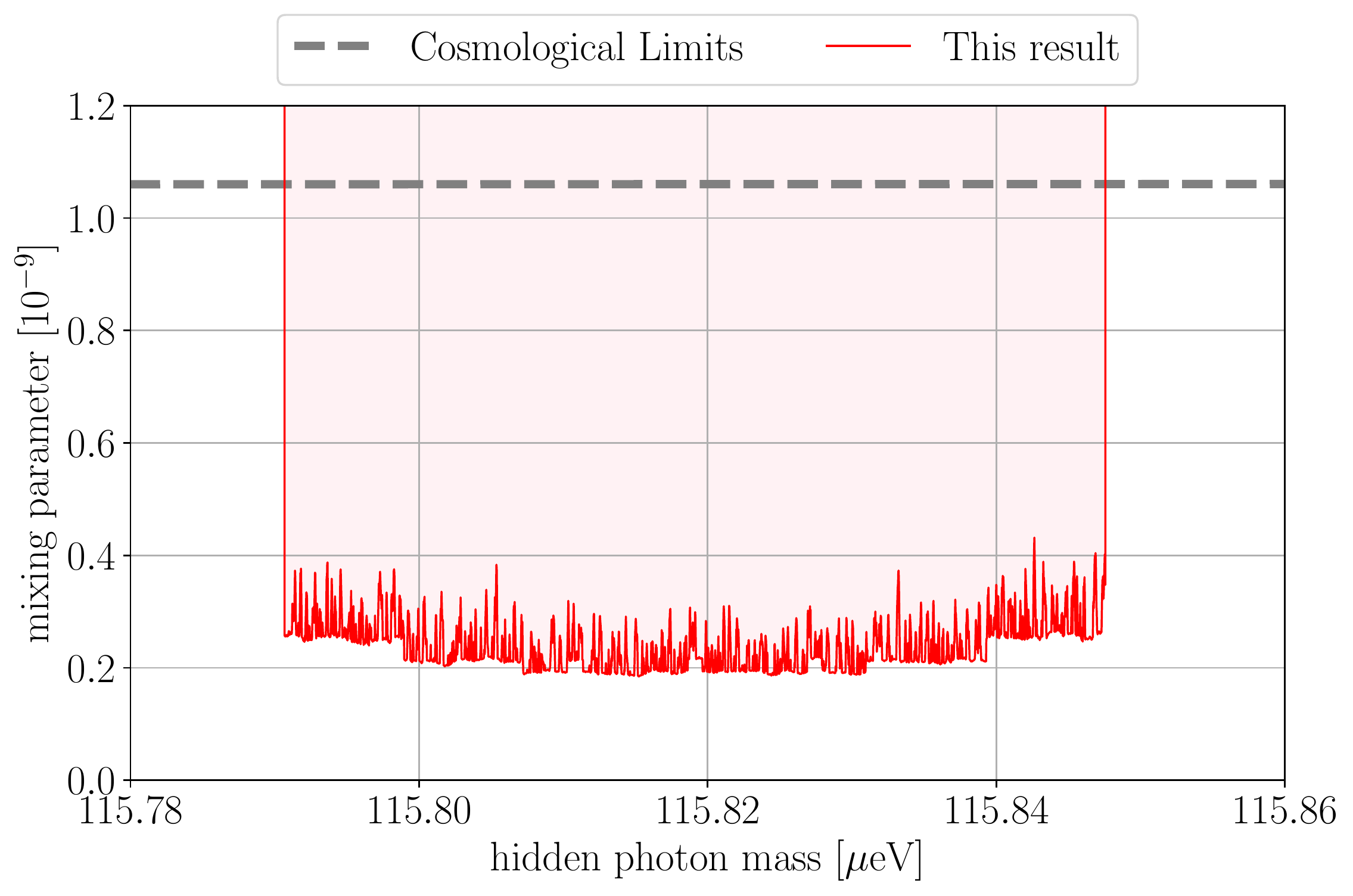}
 \caption{
 Constraints of the kinetic mixing parameter as a function of the HP-CDM mass.
 Shaded areas are the excluded parameter space.
 The bottom plot having a linear scale is an expanded view of the top plot around the mass region searched in the present paper.
 Results of previous research are also shown \cite{Arias2012, An2013190, 1475-7516-2013-08-034, Suzuki2015a, Suzuki:2015vka, Xenon10_2017, Knirck_2018, SHUKET2019, DAMIC2019, FUNK2020}.
 We set the most stringent limits in the mass range from 115.79 to 115.85\,$\mu$eV.}
 \label{fig:limits}
\end{figure}

We calculate limits of power at the 95\% confidence level for each frequency bin using 
\begin{equation}
 {\rm max}(0, P_{\gamma'}) + 1.65 \Delta P_{\gamma'}.
\end{equation}
Adopting eq.~\eqref{eq:mixing_angle}, we set upper limits for the mixing angle as shown in Figure~\ref{fig:limits}.
Obtained limits are $\chi <$ 1.8--4.3 $\times 10^{-10}$ at the 95\% confidence level.
This is the most stringent limit obtained to date in the considered mass range.

\section{Conclusions}
\label{sec:conclusion}

We searched for HP-CDM
in the mass region of 115.79--115.85 $\mu$eV
using a metal plate as the photon converter and cryogenic receiver as a photon detector in the millimeter wave range.
No excess from the zero-signal hypothesis was found in this region.
We set upper limits, $\chi <$ 1.8--4.3 $\times 10^{-10}$, at the 95\% confidence level.

\acknowledgments
This work is supported by JSPS KAKENHI under grant numbers 16K13809 and 19H05499,
JST START (Program for Creating STart-ups from Advanced Research and Technology), and the Special Postdoctoral Researchers Program in RIKEN.
We thank M.~Hasegawa and Y.~Minami for comments on the draft,
and Glenn Pennycook, MSc, from Edanz Group (https://en-author-services.edanzgroup.com/) for editing a draft of this manuscript.
\bibliographystyle{JHEP}
\bibliography{refs}

\end{document}